# Temperature-Dependent Infrared Reflectivity Studies of Multiferroic TbMnO$_3$: Evidence for Spin-Phonon Coupling

Pradeep Kumar[1], Surajit Saha[1], C R Serrao[2,3], A K Sood[1,3,*] and C N R Rao[3]

[1]Department of Physics, Indian Institute of Science, Bangalore-560012, India

[2]Material Research Centre, Indian Institute of Science, Bangalore-560012, India
[3]Chemistry and Physics of Materials Unit and International Centre for Materials Science, Jawaharlal Nehru Centre for Advanced Scientific Research, Bangalore-560064, India

*E-mail: asood@physics.iisc.ernet.in, Ph: +91-80-22932964

**Abstract**

We have measured near normal incidence far infrared (FIR) reflectivity spectra of a single crystal of TbMnO$_3$ from 10K to 300K in the spectral range of 50 cm$^{-1}$ to 700 cm$^{-1}$. Fifteen transverse optic (TO) and longitudinal optic (LO) modes are identified in the imaginary part of the dielectric function $\varepsilon_2(\omega)$ and energy loss function Im($-1/\varepsilon(\omega)$), respectively. Some of the observed phonon modes show anomalous softening below the magnetic transition temperature T$_N$ (~ 46K). We attribute this anomalous softening to the spin-phonon coupling caused by phonon modulation of the super-exchange integral between the Mn$^{3+}$ spins. The effective charge of oxygen (Z$_O$) calculated using the measured LO-TO splitting increases below T$_N$.

Keywords: Spin-phonon coupling ; Longitudinal and Transerve optic modes; Effective charge; Energy loss function.

## 1. Introduction

Multiferroics are materials in which electric and magnetic properties are correlated. These materials have attracted much attention in recent years because of their magneto-electric phenomena as well as potential applications [1-4]. Only a few systems with strong magneto-electric effect are known because proper ferroelectricity and magnetism tend to be mutually exclusive and interact very weakly whenever they coexist [5]. Much of the recent interest is in systems with improper ferroelectricity, which is due to the exchange striction in magnetically ordered states [1] and this produces strong coupling between magnetic and ferroelectric order parameters. A strong coupling between spin, charge and lattice degrees of freedom in perovskite manganite, $RMnO_3$ (R = Rare Earth), gives many competing phases. An identification of large magneto-electric and magnetocapacitive effects in undoped perovskite manganites with small sized R cations, e.g., $TbMnO_3$ and $DyMnO_3$, has made the family of manganites even more interesting [6,7]. It has been shown that an external electric field can change the magnetic domain distribution [8]. Depending on the size of the R-ion, these materials have two kinds of crystal structure. $RMnO_3$ crystallize in orthorhombic structure for larger ionic radius (R = La, Ce, Pr, Nd, Sm, Eu, Gd, Tb and Dy) and in hexagonal structure for R with smaller ionic radius (R = Ho, Er, Tm, Yb, Lu, Y). All the $RMnO_3$ perovskites show a distortion of $MnO_6$ octahedra due to Jahn-Teller (JT) effect of $Mn^{3+}$ cations [9-11]. At room temperature $RMnO_3$ shows paramagnetic and insulating behavior. The latter arises from splitting of the $e_g$ level of the d-electrons of the Mn-ions, caused by the JT distortion of the $MnO_6$ octahedra.

$TbMnO_3$ is orthorhombic (space group Pbnm) at room temperature and shows an incommensurate lattice modulation at $T_N$ for sinusoidal antiferromagnetic ordering, with $T_N \sim$ 41K [6] or $T_N \sim$ 46K as reported by Bastjan et al. [12]. Ferroelectric order develops at the income mensurate-commensurate transition temperature $T_{FE} \sim$ 27K [6]. The lattice modulation below $T_N$ is caused by the $Mn^{3+}$ ions increasing their exchange interaction energy by shifting their positions [13, 14]. It has been suggested that the origin of ferroelectricity in $TbMnO_3$ is induced by the complex spin structure [7]. As the temperature is further lowered, rare-earth $Tb^{3+}$ ions also order anti-ferromagnetically in $TbMnO_3$ at T $\sim$ 7K [6].

The role of phonons in manganites has been studied in recent years using Raman and infrared spectroscopies. In recent years, focus has been on the electromagnons observed below 150 cm$^{-1}$ [15-19]. Electromagnons are the spin waves that are excited by an ac electric field. These electromagnons are qualitatively different from the spin-phonon coupling. It has been shown that in $RMnO_3$ (R = La, Nd, Sm, Gd, Dy, and Pr), a few Raman phonons involving oxygen vibrations are anomalous i.e. the phonon frequency decreases as temperature is lowered below $T_N$ [20-22] attributed to spin-phonon coupling. However, there have not been detailed temperature dependent studies on infrared (IR) phonons in RMnO$_3$, except the work of Paolone et al. [23] wherein they have reported IR mode frequencies at 300K and 10K for undoped and doped LaMnO$_3$. From their tabulated data, it is seen that three modes in undoped and one mode in doped samples show lower frequencies at 10K as compared to their values at 300K. In this paper, we report a detailed temperature dependent study of infrared phonons in TbMnO$_3$ in the 10K to 300K range.

## 2. Experimental details

A floating-zone furnace fitted with two ellipsoid halogen lamps (NEC-Japan) with radiation heating was used to grow single crystals of TbMnO$_3$. Polycrystalline rods (feed and seed) were prepared by conventional solid-state reaction method. Stoichiometric mixtures of the starting materials Tb$_2$O$_3$ and MnO$_2$ were weighed in the desired proportions and grounded for a few hours in an agate mortar with propanol. The powder is heated at 1173 K, 1273 K, 1473 K with intermediate grinding. The powder was finally sintered at 1673 K for 24 h in air. The sample was then regrounded and monophasic polycrystalline powder was hydrostatically pressed and sintered at 1473 K for 24 h in air to obtain the feed and seed rods with a diameter of 4 mm and a length of 100 mm. A single crystal was then grown under an air atmosphere at a growth rate of 6 mm/h.

Reflectance spectra of the unoriented TbMnO$_3$ single crystal were obtained in near normal incidence geometry using a Fourier transform IR spectrometer (Bruker IFS 66v/s) in the frequency range of 50 to 700 cm$^{-1}$. The noise level below 150 cm$^{-1}$ did not allow us to study electromagnons in our experiment. We will focus on the temperature dependence of the infrared active phonons with frequencies above 150 cm$^{-1}$. Spectra were collected with an instrumental resolution of 2 cm$^{-1}$. The background reference signal was collected using a gold plated mirror. Temperature dependent reflectance spectra at temperatures ranging from 10K to 300K were collected by mounting the sample on the cold finger of a continuous flow liquid helium cryostat (Optistat CF-V, Oxford Instruments) and the temperature was controlled to an accuracy of ± 0.1K by a temperature controller (Oxford Instruments).

## 3. Results and Discussion

The temperature dependent reflectivity of TbMnO$_3$ has been shown in figure 1 for a few typical temperatures. The bands in the reflectivity spectra are due to IR active phonon modes. The total numbers of IR active phonon modes of TbMnO$_3$ (space group Pbnm), calculated using group theory are 25 and these are classified as $\Gamma_{IR}$ = 9B$_{1u}$ + 7B$_{2u}$ + 9B$_{3u}$ [22]. The eigenvectors of these modes have been reported [24] using rigid ion model. The dielectric function, $\varepsilon(\omega)$, was obtained by Kramer-Kronig (KK) analysis of the reflectance spectra using the OPUS software (Bruker Optics). The extrapolation procedure used in KK analysis was as follows: reflectivity value R ($\omega$) below 50 cm$^{-1}$ is same as R (50 cm$^{-1}$) and R ($\omega > 700$ cm$^{-1}$) = R ($\omega = 700$ cm$^{-1}$). Peak positions in the imaginary part of the dielectric function ($\varepsilon_2(\omega)$) and energy loss function (Im (-1/$\varepsilon(\omega)$)) give the frequencies of the transverse optic (TO) and longitudinal optic (LO) modes, respectively. Temperature dependent $\varepsilon_2(\omega)$ shown in figure 2 for a few typical temperatures clearly reveal fifteen TO modes, labeled as T1 to T15, apart from some weak modes which appear as shoulders (not labeled). It is likely that the band labeled as T3 can be a combination of two modes, as indicated by its lineshape at low temperatures. The corresponding LO modes are identified as L1 to L15 in Im(-1/$\varepsilon(\omega)$), shown in figure 3 for a few typical temperatures. We tried to fit $\varepsilon_2(\omega)$ and Im(-1/$\varepsilon(\omega)$) by a sum of fifteen Lorentizian like functions [25, 26] to extract the peak positions, linewidth and oscillator strengths of the IR modes. This could not be done due to large number of parameters (15×3 + one for $\varepsilon(\infty)$ + base line parameter = 47). We, therefore, have extracted only the peak positions by directly reading from the spectra of $\varepsilon_2(\omega)$ and Im(-1/$\varepsilon(\omega)$). These values are given in table 1 for 300K to give an indication of the LO-TO splitting in TbMnO$_3$. The mode frequencies are close to the observed [23] and calculated [24] frequencies for the infrared phonons of LaMnO$_3$.

We will now discuss the temperature dependence of the LO and TO modes extracted from the reflectivity measurements. It is seen that the temperature dependence of nine modes (modes 1-3, 7, 9, 10, 11, 14 and 15) is very weak and normal (data not shown) i.e. the mode frequency increases by less than 1 or 2 cm$^{-1}$ as temperature is lowered to 10K. Figure 4 and figure 5 show the temperature dependence of other modes 4, 5, 6, 8, 12 and 13 for both the LO and TO components which reveal significant temperature dependence.

In general, the temperature dependent behavior of a phonon mode of frequency '$\omega$' is given as [21]

$$\omega(T) = \omega(0) + (\Delta\omega)_{qh}(T) + (\Delta\omega)_{anh}(T) + (\Delta\omega)_{el\text{-}ph}(T) + (\Delta\omega)_{sp\text{-}ph}(T) \qquad (1)$$

where $\omega(0)$ is the phonon frequency at T = 0K. The term "$(\Delta\omega)_{qh}(T)$" corresponds to the change in phonon frequency due to a change in the unit cell volume, termed as "quasiharmonic" effect. $\Delta\omega_{anh}(T)$ represents the intrinsic anharmonic contributions to the phonon frequency, which is related to the real part of the phonon self-energy. The effect of renormalization of the phonon frequency (($\Delta\omega)_{el\text{-}ph}(T)$) due to electron-phonon coupling is absent in insulating

TbMnO$_3$. The term $\Delta\omega_{sp\text{-}ph}$(T) is the change in phonon frequency due to spin-phonon coupling, caused by the modulation of the exchange integral by lattice vibrations [21]. The change in phonon frequency of mode "$i$" due to change in lattice constant, i.e ($\Delta\omega$)$_{qh}$(T), can be related to the change in volume if we know Grüneisen parameter $\gamma_i = -(B_0/\omega_i)(\partial\omega_i/\partial P)$, where $B_0$ is the bulk modulus and $\partial\omega_i/\partial P$ is the pressure-derivative of the phonon frequency. For a cubic crystal or isotropically expanded lattice, the change in phonon frequency due to change in volume is given as $(\Delta\omega)_i(T)_{qh}/\omega_i(0) = -\gamma_i(\Delta V(T)/V(0))$. In the Raman study of RMnO$_3$ [R = Gd, Eu, Pr, Nd, Sm, Tb, Dy, Ho, and Y], the quasi harmonic contribution has been neglected [20, 22]. Similarly, the quasi harmonic contribution in TbMnO$_3$ can also be neglected because coefficient of thermal expansion for lattice parameter c remain nearly constant whereas for a and b it varies in opposite direction [27]. Therefore, fractional change in volume is negligible and hence $(\Delta\omega_i)_{qh}$(T) can be ignored.

In a cubic anharmonic process, a phonon of frequency $\omega(\vec{\kappa}=0)$ decays into two phonons $\omega_1(\vec{\kappa_1})$ and $\omega_2(\vec{\kappa_2})$, keeping energy and momentum conserved, i.e. $\omega = \omega_1 + \omega_2$, $\vec{\kappa_1} + \vec{\kappa_2} = 0$. Considering the simplest decay channel with $\omega_1 = \omega_2$, the temperature dependence of ω (T) can be expressed as [28]

$$\omega(T) = \omega(0) + C\,[1+2n(\omega(0)/2)] \qquad (2)$$

where $\omega$(T) is the phonon frequency at temperature T, $\omega$(0) is the phonon frequency at T = 0K in the harmonic approximation, C is self-energy parameter for a given phonon mode and $n(\omega) = 1/(\exp(\hbar\omega/\kappa_B T) -1)$ is the Bose–Einstein mean occupation number. We have fitted the modes T4, T5, T6; L4, L5, L6 and L13 using equation (2). The solid lines, in figure 4 and figure 5, correspond to the fits by equation (2), with fitting parameters given in table 1.

The most interesting observation from our experiments is that the modes T8, T12, T13 (figure 4) and L12 (figure 5) show anomalous temperature dependence: the modes show softening below T$_N$. Similar anomalous temperature dependence has been observed for a few Raman modes in RMnO$_3$ where R = La [16] and Gd, Pr, Nd, Sm, Dy [20, 22], which has been attributed to spin-phonon coupling [21]. This is understood as follows: if an ion is displaced from its equilibrium position by "$u$", then the crystal potential is given as $U = (1/2) * (ku^2) + \Sigma_{ij} J_{ij}(u) S_i S_j$, where k in the first term represent the force constant and the second term arises from spin interactions between the $Mn^{3+}$ spins; second derivative of the crystal potential (U) gives a harmonic force constant. The phonon frequency is affected by the additional term i.e. $(\Delta\omega)_{sp-ph}(T) = \lambda <S_iS_j>$, where $\lambda = (\partial^2 J_{ij}(u)/\partial u^2)$ is the spin-phonon coupling coefficient and $<S_iS_j>$ is the spin-correlation function. $\lambda$ can be positive or negative and is different for different phonons. As the temperature decreases, the spin correlations build up and hence the spin-phonon coupling becomes important at lower temperatures. Therefore, renormalization of the phonon frequency below $T_N$ is expected in $TbMnO_3$. We note that the phonon softening for the T13 mode is similar in magnitude to the $B_{2g}$ (~ 604 $cm^{-1}$) mode in $LaMnO_3$ [21] and hence the spin-phonon coupling constant ($\lambda$) is expected to be similar for both the modes.

We now comment on the temperature dependence of the frequency difference between the LO-TO modes which determines the effective charge of the ions in the lattice. The effective charge of Tb, Mn and O ions in the unit cell with k atoms can be estimated using the relation [29]:

$$\Sigma_k(Z^2_k/M_k) = (\pi V) * \Sigma_j(\omega^2_{LO,j} - \omega^2_{TO,j}) \qquad (3)$$

where V is the unit cell volume, j is an index for the lattice modes. $Z_k$ is the effective charge of the $k^{th}$ ion with mass $M_k$. For the effective charges, there is a sum rule for charge neutrality i.e $\Sigma_k (Z_k) = 0$. We note that the effective charge calculated using the above equation is not same as the ionic charge, the Szigeti charge or the nominal valence charges [30]. In $TbMnO_3$, where the masses of other constituents are much heavier than the oxygen mass, we can neglect the terms other than the terms for oxygen on the left hand side of equation (3). The calculated value of the effective charge of oxygen from equation (3) is the average for all the oxygen sites. Using the observed values of the TO and the LO frequencies in the equation (3) the temperature dependence of the oxygen effective charge ($Z_O$) is calculated, as shown in figure 6. The absolute value of $Z_O$ is not crucial, what is more important is its change with temperature. We find that $Z_O$ increases below $T_N$. If $Z_O$ increases, then the induced dipole moment and hence the optical absorption will also increase. The increase in $Z_O$ below $T_N$ suggests a change in the bond length between $Mn^{3+}$ cations and the corresponding bond angles, mediated by the O-ions. This result needs to be understood better.

In conclusion, infrared reflectivity measurements of a single crystal of $TbMnO_3$ clearly identify fifteen IR active phonon modes. Out of these, three modes show anomalous behavior below the magnetic phase transition temperature $T_N$ which is attributed to spin-phonon coupling. The effective charge of the oxygen ions shows an increase below $T_N$. We hope that our results on phonon softening and an increase in the effective charge of oxygen ions will motivate theoretical studies of phonons in TbMnO$_3$ and their role in the magneto-electric behavior.

**Acknowledgments**
AKS thanks the Department of Science and Technology (DST), India, for financial support. Pradeep Kumar thanks Council of Scientific and Industrial Research (CSIR), India, for research fellowship.

**References**

[1] Y Tokura   *Science* **312,** 1481 (2006)

[2] W Eerenstein, N D Mathur and J F Scott   *Nature* **442,** 759 (2006)

[3] C N R Rao and C R Serrao   *J. Mater. Chem* **17,** 4931(2007)

[4] S W Cheong and M Mostovy   *Nature Mater.* **6,** 13 (2007)

[5] N A Hill and A Filippetti J. *Magn. Magn. Mater.* **976,** 242 (2002)

[6] T Kimura , T Goto , H Shintani , K Ishizaka , T Arima  and  Y Tokura   *Nature* **426,**  55 (2003)

[7] T Goto, T Kimura, G Lawes, A P Ramirez  and Y Tokura   *Phys. Rev. Lett.* **92,** 257201 (2004)

[8] Y Yamasaki, H Sagayama, T Goto,  M Matsuura, K Hirada and Y Tokura Phys. *Rev. Lett.* **98,** 147204 (2007)

[9] L M Carron , A de Andres , M J Martinez Lope , M T Casais  and J A  Alonso  *Phys. Rev.*  B **66,** 174303 (2002)

[10] W Wei-Ran, X D Peng and S W Hui Chin. *Phys. Lett.* **22,** 705 (2005)

 [11] M N Iliev ,  M V Abrashev ,  J Laverdiere , S  Jandl , M M  Gospodinov , Y Q Wang  and  Y Y Sun   *Phys. Rev.* B **73,** 064302 (2006)

[12] M Bastjan , S G Singer , G Neuber , S Eller ,  N Aliouane , D N Argyrion , S L Cooper  and M Rubhausen   *Phys. Rev.* B **77,**  193105 (2008)

[13] S Greenwald and J S Smart Nature **166,** 523 (1950)

[14] J S Smart and S Greenwald   *Phys. Rev.* **82,** 113 (1951)

[15] A Pimenov , A A Mukhin , V Yu Ivanov , V D Travkin , A M Balbashov and A Loidl *Nature Physics* **2**, 97 (2006)

[16] R V Aguilar , A B Sushkov , C L Zhang , Y J Choi , S W Cheong and H D Drew *Phys. Rev. B* **76**, 060404 (2007)

[17] Y Takahashi , N Kida , Y Yamasaki , J Fujioka , T Arima , R Shimano , S Miyahara , M Mochizuki , N Furukawa and Y Tokura *Phys. Rev. Lett.* **101**, 187201 (2008)

[18] R V Aguilar, M Mostovoy, A B Sushkov , C L Zhang , Y J Choi , S W Cheong and H D Drew *Phys. Rev. Lett.* **102**, 047203 (2009)

[19] A Pimenov , A Shuvaev, A Loidl , F Schrettle , A A Mukhin , V D Travkin , V Yu Ivanov and A M Balbashov *Phys. Rev. Lett*. **102**, 107203 (2009)

[20] J Laverdiere, S Jandl, A A Mukhin, V Yu Iyanov, V G Iyanov and M N Iliev *Phys. Rev.* B **73,** 214301 (2006)

[21] E Granado , A Garcia , J A Sanjurjo , C Rettori and I Torriani *Phys. Rev.* B **60,** 11879 (1999)

[22] W S Ferreira , J A Moreira , A Almeida , M R Chaves , J P Araujo , J B Oliveira , J M Machado Da Silva , T M Sa , T M Mendonca and P S Carvalho *Phys. Rev.* B **79,** 054303 (2009)

[23] A Paolone, P Roy, A Pimenov, A Loidl, O K Melnikev and A Ya Shapiro *Phys. Rev.* B **61,** 11255 (2000)

[24] I S Smirnova *Physica* B **262,** 247 (1999)

[25] C C Holmes, Vogt T, Shapiro S M, Wakimoto S, M A Subramanian and A P Ramirez *Phys. Rev.* B **67** 092106 (2003)

[26] S Tajima , T Ido , S Ishibashi , T Itoh , H Eisaki , Y Mizauo , T Arima , H Takagi and S Uchido *Phys. Rev.* B **43,** 10496 (1991)

[27] D Meier, N Aliouane , D N Argyriou , J A Mydosh and T Lorenz *New J. Phys.* **9,** 100 (2007)

[28] P G Klemens *Phys. Rev.* **148,** 845 (1966)

[29] J F Scott *Phys. Rev.* B **4,** 1360 (1971)

[30] W Cochran *Nature* **191,** 60 (1961)

**Table1.** List of the experimental phonon frequencies at 300K and fitting parameters of a few phonons, fitted by equation (2) as described in text. Units are in cm$^{-1}$.

| Mode label Number | TO mode frequency | | | LO mode frequency | | |
|---|---|---|---|---|---|---|
| | $\omega(300)$ | $\omega(0)$ | C | $\omega(300)$ | $\omega(0)$ | C |
| 1. | 176 | | | 177 | | |
| 2. | 197 | | | 201 | | |
| 3. | 252 | | | 260 | | |
| 4. | 291 | 298.2 ± 0.9 | -3.4 ± 1.4 | 295 | 300.6 ± 0.4 | -3.1 ± 1.2 |
| 5. | 309 | 312.4 ± 1.1 | -2.7 ± 0.6 | 309 | 313.6 ± 1.2 | -2.1 ± 0.8 |
| 6. | 335 | 339.6 ± 1.3 | -3.6 ± 1.4 | 338 | 342.6 ± 0.4 | -3.5 ± 0.8 |
| 7. | 384 | | | 387 | | |
| 8. | 419 | | | 425 | | |
| 9. | 446 | | | 449 | | |
| 10. | 465 | | | 474 | | |
| 11. | 511 | | | 520 | | |
| 12. | 552 | | | 555 | | |
| 13. | 601 | | | 614 | 623.2 ± 1.1 | - 3.6 ± 0.9 |
| 14. | 641 | | | 657 | | |
| 15. | 686 | | | 697 | | |

**Figure caption:**

**Fig.1** Temperature dependent infrared reflectivity of TbMnO$_3$ in the spectral range 150 cm$^{-1}$ to 700 cm$^{-1}$.

**Fig.2** Temperature dependence of the imaginary part of the dielectric function ($\varepsilon_2(\omega)$).

**Fig.3** Evolution of the energy loss function ( Im(-1/$\varepsilon(\omega)$) ) with temperature.

**Fig.4** Temperature dependence of the TO modes T4, T5, T6, T8, T12 and T13. Solid lines for T4, T5 and T6 are the fitted curves as described in the text. The frequencies for the T8, T12 and T13 modes have been fitted by a linear relation above T$_N$.

**Fig.5** Temperature evolution of the LO modes L4, L5, L6, L8, L12 and L13. Solid lines for L4, L5, L6 and L13 are the fitted curves as described in the text.

**Fig.6** Temperature dependence of the effective charge of oxygen (Z$_O$). Solid line shows a linear fit in the temperature range above T$_N$.

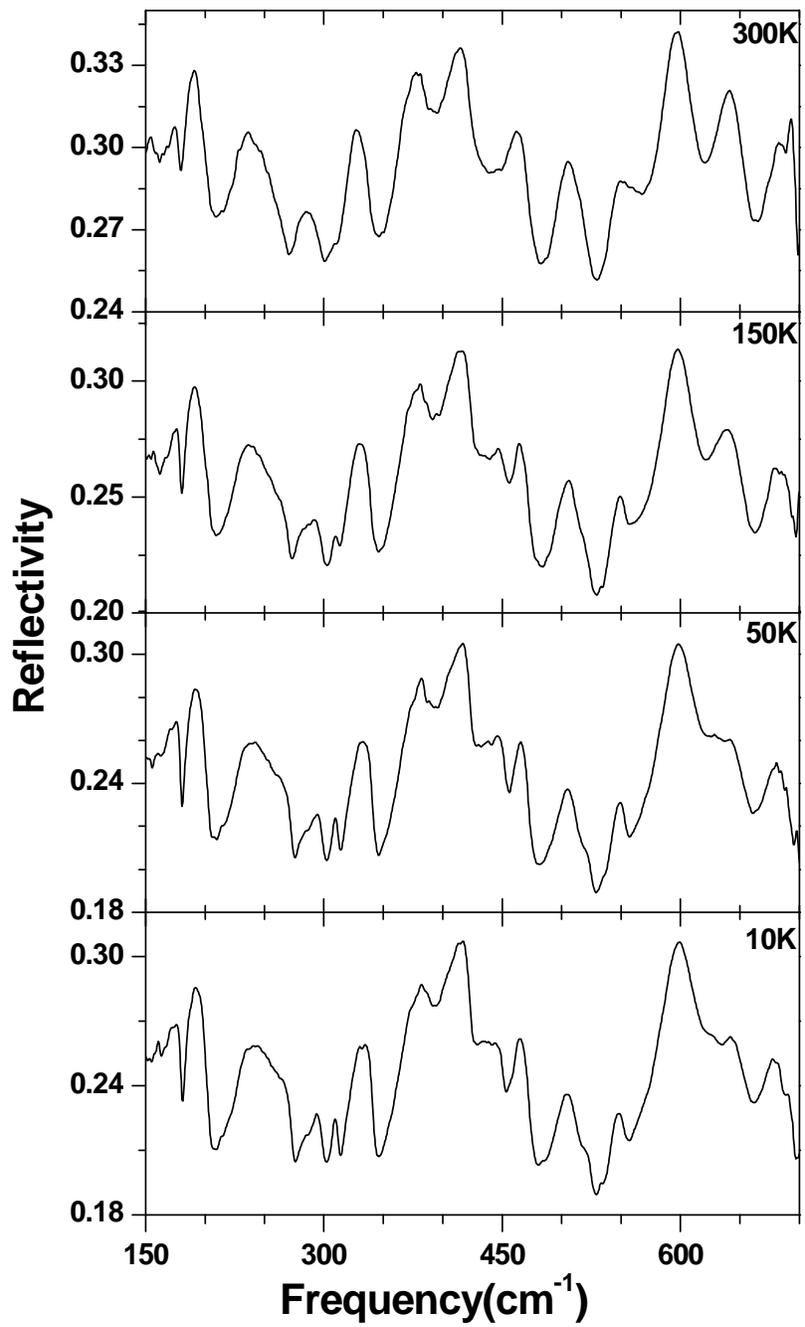

Figure1.

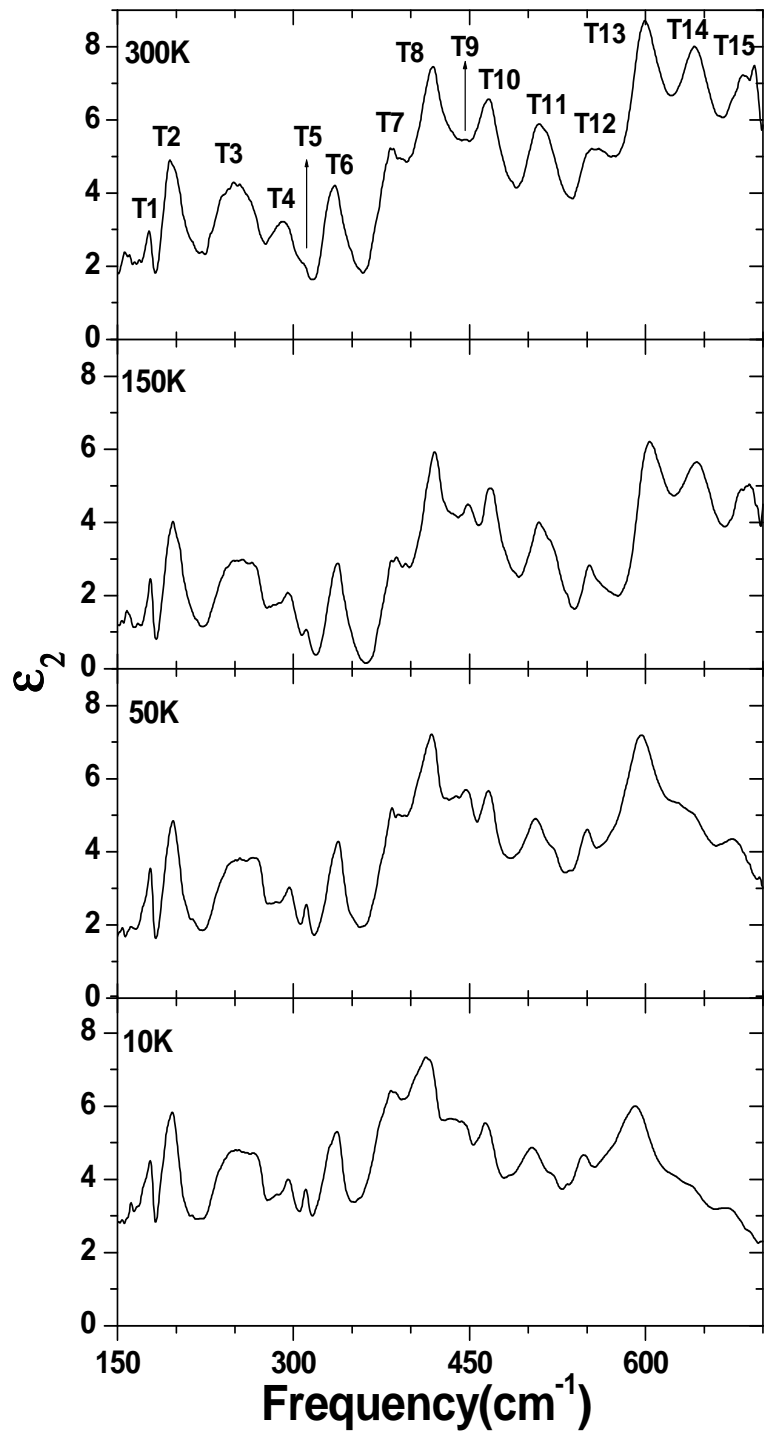

**Figure 2.**

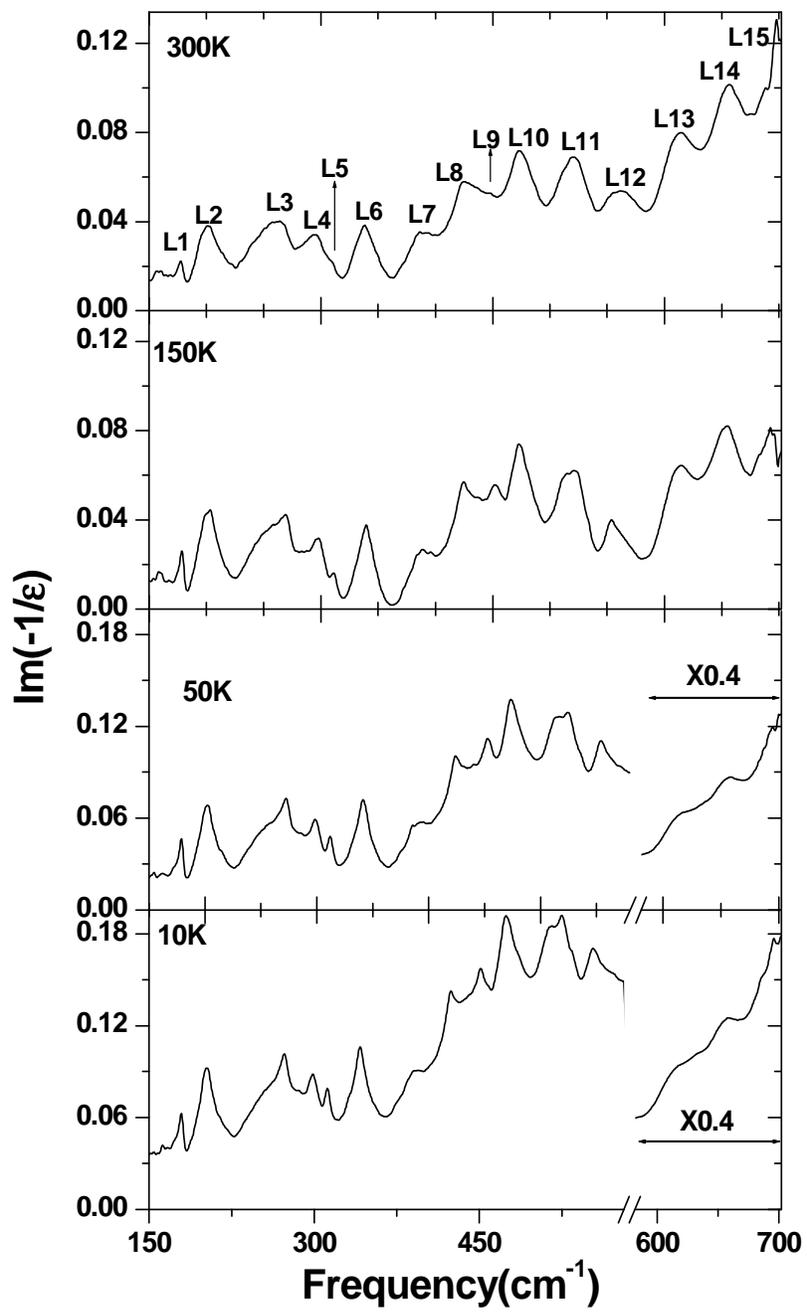

Figure3.

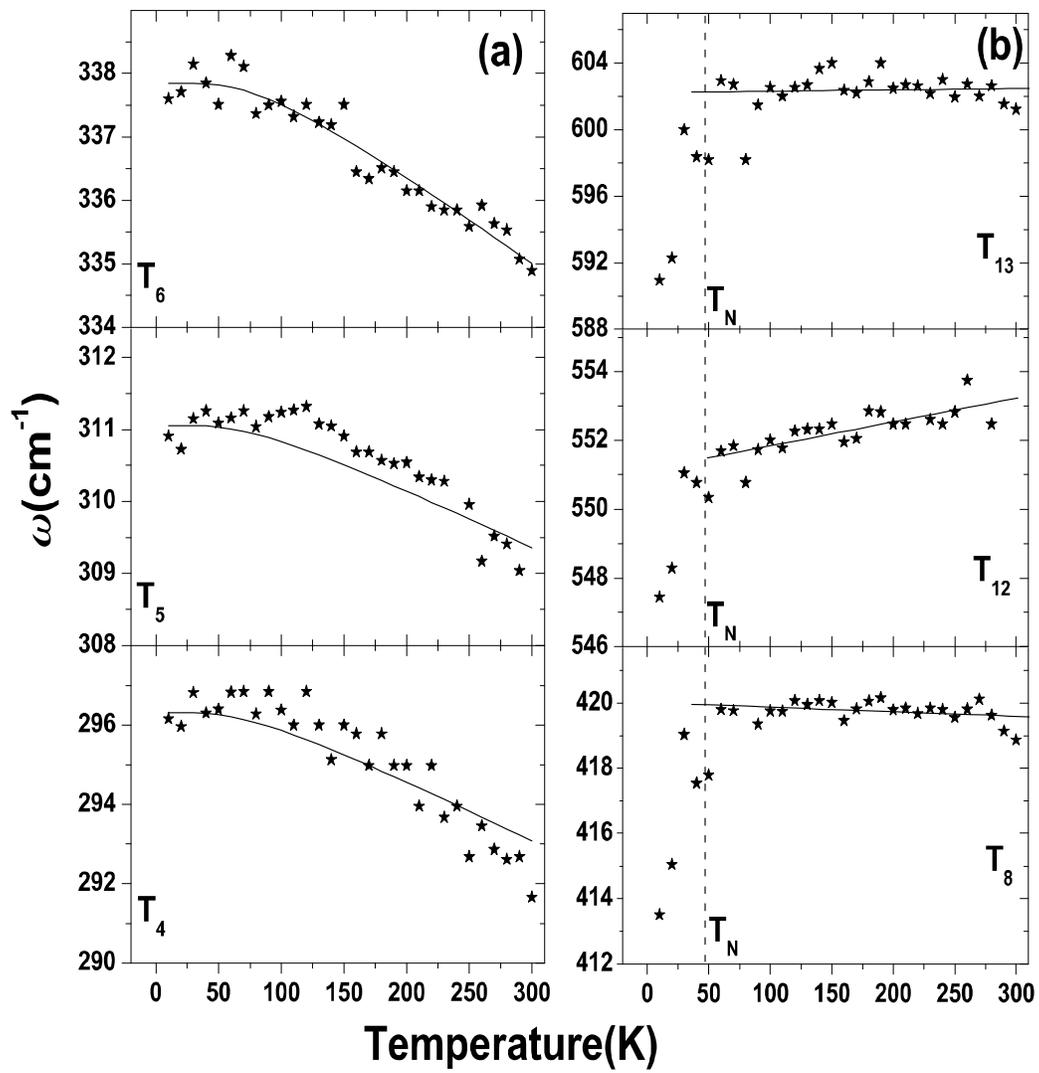

Figure 4.

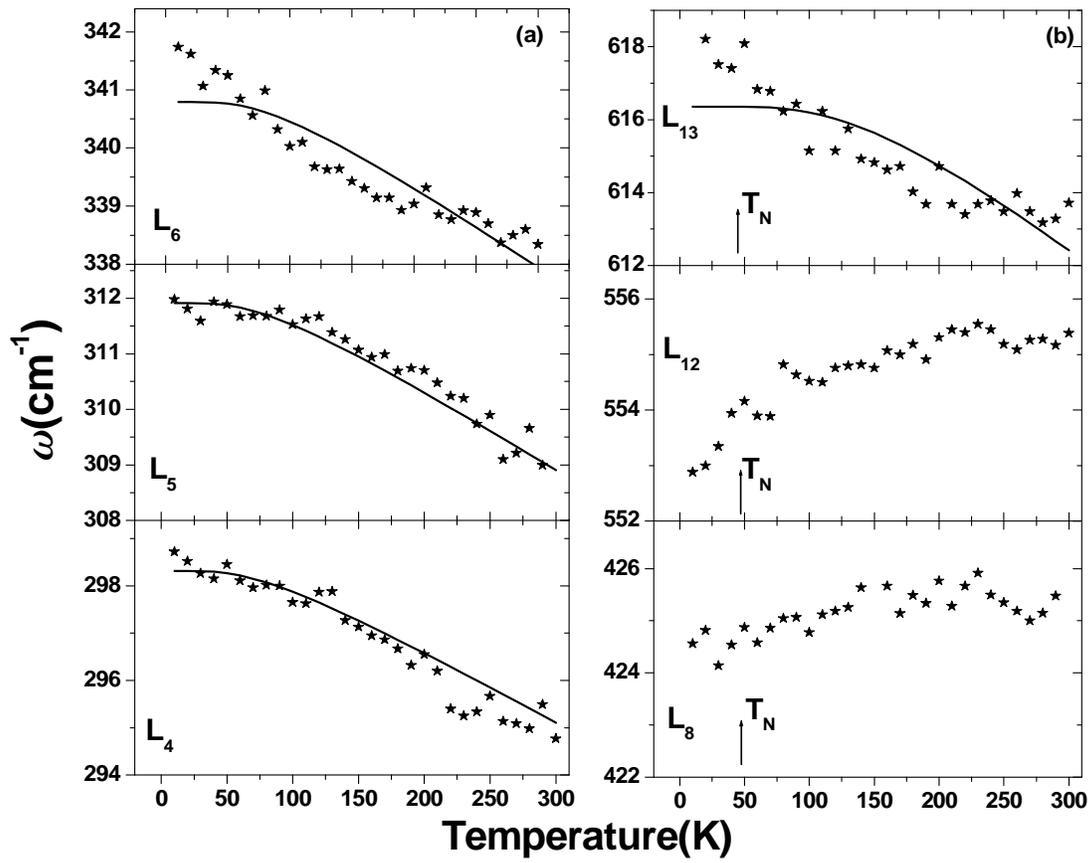

**Figure5.**

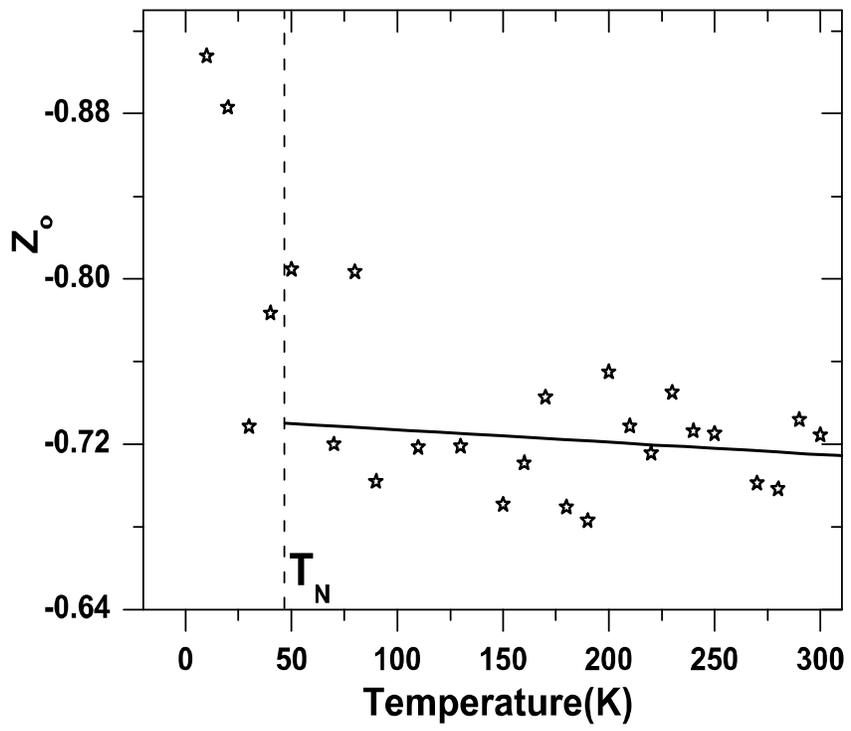

**Figure 6.**